\documentclass[preprint,12pt]{elsarticle}

\usepackage{amssymb}
\usepackage{amsmath}

\newcommand{\xg}[1]{#1}

\journal{Nuclear Instruments and Methods in Physics Research Section A}

\begin{document}

\begin{frontmatter}



\title{A water Cherenkov detector prototype for future high-energy tau-neutrino experiment}


\author[1]{Monong Yu\corref{cor1}} 
\ead{mnyu@pmo.ac.cn}  
\author[1,2]{Yiren Chen}
\author[1,2]{Lei Wang}
\author[1]{Shiping Zhao\corref{cor1}}
\ead{zhaosp@pmo.ac.cn}  
\author[1,2]{Xingyu Li}
\author[1]{Qiang Yuan}
\author[1]{Yi Zhang}
\author[3]{Guang Yang} 

\affiliation[1]{organization={Purple Mountain Observatory, Chinese Academy of Science},
            city={Nanjing},
            postcode={210023}, 
            state={Jiangsu},
            country={China}}
\affiliation[2]{organization={University of Science and Technology of China},
            city={Hefei},
            postcode={230026}, 
            state={Anhui},
            country={China}}
\affiliation[3]{organization={Nanjing Institute of Astronomical Optics and Technology},
            city={Nanjing},
            postcode={210042}, 
            state={Jiangsu},
            country={China}}

\cortext[cor1]{Corresponding author.}

\begin{abstract}
The detection of high-energy tau neutrinos ($\nu_{\tau}$) remains a critical challenge in neutrino astronomy, limited by inadequate angular resolution and sensitivity in current detectors like IceCube and KM3NeT. We present a modular water Cherenkov detector prototype optimized for $\nu_{\tau}$-induced extensive air showers (EAS) in the 1–100 PeV range, leveraging canyon terrain for natural cosmic-ray shielding. Laboratory validation demonstrates this prototype design has high detection efficiency (\textgreater99\%) and timing resolution (\textless2 ns) on MIP particles, enabling precise $\nu_{\tau}$-induced EAS reconstruction for future study. The results establish a foundation of a low-cost, scalable neutrino observatory, advancing flavor ratio measurements and cosmic-ray origin problems.
\end{abstract}




\end{frontmatter}



\section{Introduction}
\label{sec1}
Neutrino astronomy has revolutionized our understanding of high-energy astrophysical processes, with landmark discoveries from IceCube\cite{achterberg2006first} firmly establishing astrophysical neutrinos as indispensable cosmic messengers \cite{icecube2018multimessenger}. By detecting neutrinos from extragalactic sources such as active galactic nuclei (AGN) and gamma-ray bursts (GRBs), IceCube has unveiled a non-electromagnetic window into the most extreme environments in the universe \cite{icecube2013evidence,albert2017search}. The success of IceCube has spurred the development of next-generation neutrino telescopes, \xg{including IceCube-Gen2 \cite{aartsen2021icecube}, an expansion of IceCube}, KM3NeT in the Mediterranean \cite{adrian2016letter,kopper2012performance,km3net2025observation}, Baikal-GVD in Siberia \cite{allakhverdyan2023diffuse}, besides HUNT \cite{huang2023proposal} and TRIDENT \cite{ye2023multi} in the South China Sea, each aiming to expand detection sensitivity across energy ranges and celestial regions. These efforts highlight the critical role of neutrinos in probing cosmic-ray acceleration mechanisms and extreme astrophysical phenomena unobservable through traditional electromagnetic ways.

Despite these advancements, a glaring gap persists in the detection of high-energy tau-neutrinos ($\nu_{\tau}$). \xg{Existing detector designs have remained confined to the sub-PeV regime and predominantly focus on $\nu_{\mu}$ and $\nu_{e}$ flavors due to technical constraints. The flavor bias and the energy-coverage gap arise the need of novel detection technologies to access the elusive $\nu_{\tau}$ channel and extend the capacity to 100 PeV to explore fundamental astrophysical processes such as Greisen-Zatsepin-Kuzmin neutrino production.} While IceCube has reported tentative $\nu_{\tau}$ candidates \cite{abbasi2024observation,aartsen2016search}, its angular resolution (\textgreater5°) and effective volume in the ice medium remain inadequate for definitive $\nu_{\tau}$ identification or source localization. Moreover, deep-water detectors like KM3NeT face prohibitive construction costs and environmental challenges, while radio-based arrays such as GRAND \cite{fang2017giant} remain experimental due to immature self-triggering techniques. This technical impasse hinders the exploration of $\nu_{\tau}$-specific physics, including flavor ratio measurements to test neutrino oscillations over astrophysical distances and the validation of cosmic-ray proton acceleration models through $\nu_{\tau}$-source correlations.

\begin{figure}
    \centering
    \includegraphics[width=300pt]{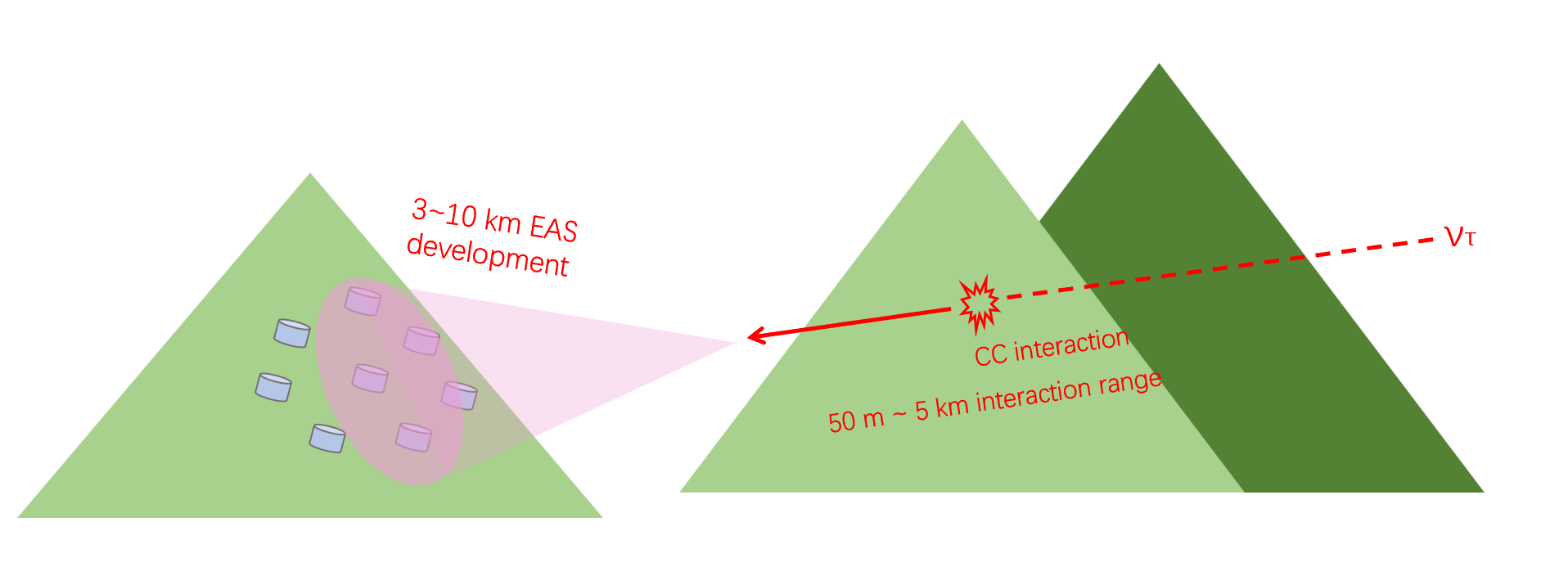}
    \caption{Schematic diagram of the proposed water Cherenkov detector array deployed in a canyon terrain, leveraging natural rock shielding and modular units for large-zenith-angle tau-induced air shower detection.}
    \label{fig:valley}
\end{figure}

To address this challenge, novel detection concepts tailored for $\nu_{\tau}$—such as the Tau Air Shower Mountain-Based Observatory (TAMBO)—have emerged \cite{2025TAMBO}. \xg{TAMBO achieves unmatched sensitivity to the diffuse astrophysical neutrino flux in the PeV-EeV energy range, surpassing all existing observatories and extending IceCube’s measurements to ultra-high energies. High-density rock enhances $\nu_{\tau}$ charged-current (CC) interaction probability, while decay process dominates $\tau$-lepton energy loss at PeV-EeV scales, during which $\nu_{\tau}$ regeneration sustains flux integrity. Simulations confirm $\tau$-exit-probabilities of 10$^{-4}$-10$^{-2}$ (depending on emergence angles), readily detectable by large-scale arrays \cite{2018comprehensive}.} Building on TAMBO's pioneering use of natural topography, we propose a cost-effective, large-area water Cherenkov detector array leveraging China’s unique canyon terrain, as shown in Fig.~\ref{fig:valley}. The natural rock walls of canyons provide inherent shielding against cosmic-ray background, while the geographic topology allows cost-efficient deployment of modular detector units across slopes optimized for capturing upward-going tau-induced extensive air showers (EAS). \xg{This design targets the 1–100 PeV energy range, where $\nu_{\tau}$ interactions exhibit enhanced signal-to-noise ratios (SNR) against atmospheric backgrounds, and exploits the distinct ``double-bang" signatures of $\tau$-lepton-decay \cite{cowen2007tau}.} Compared to ice- or deep-water-based detectors, this approach minimizes logistical complexity and environmental impact, offering a scalable and adaptable platform for $\nu_{\tau}$ astronomy. The proposed water Cherenkov detector demonstrates several advantages critical for $\nu_{\tau}$ detection: (1) A modular design using standardized units enables rapid deployment and future expansion at lower cost than IceCube’s infrastructure; (2) Advanced time synchronization via White Rabbit protocols ensures sub-nanosecond timing resolution for precise EAS reconstruction; (3) High quantum efficiency (QE) photomultipliers and machine learning-based noise suppression algorithms enhance signal discrimination in high-background environments. These innovations collectively achieve angular resolution below 1°, resolving $\nu_{\tau}$ sources with unprecedented accuracy.

In this article, we present the design, prototyping, and performance validation of a water Cherenkov detector optimized for $\nu_{\tau}$ detection. Through laboratory experiments and Monte Carlo simulations, we demonstrate its capability to isolate $\nu_{\tau}$ signals with high sensitivity and angular precision. The results confirm the prototype’s potential to form modular detector arrays in canyon environments, bridging the critical gap in $\nu_{\tau}$ astronomy while paving the way for scalable, next-generation neutrino observatories.

\section{Water Cherenkov prototype detector design}

\subsection{Detector prototype design}
\begin{figure}
    \centering
    \includegraphics[width=200pt]{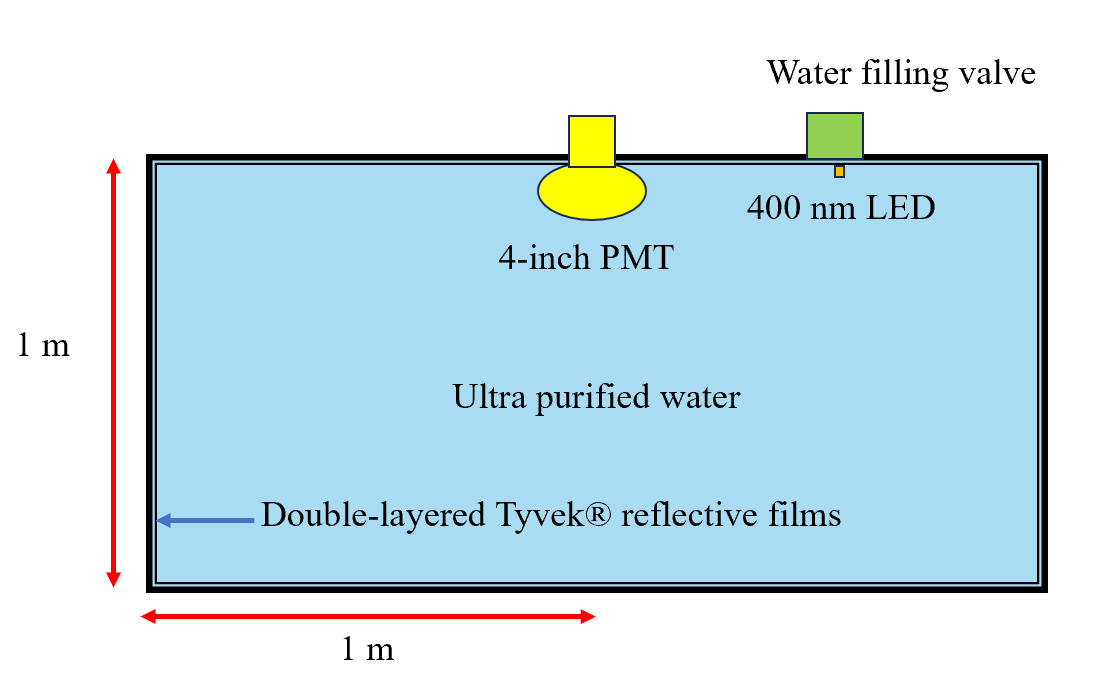}
    \caption{Cross-sectional view of the cylindrical water Cherenkov detector prototype (1 m radius, 1 m height) with Tyvek® reflective lining, central PMT placement, and LED calibration system.}
    \label{fig:tank_design}
\end{figure}

\begin{figure}
    \centering
    \includegraphics[width=200pt]{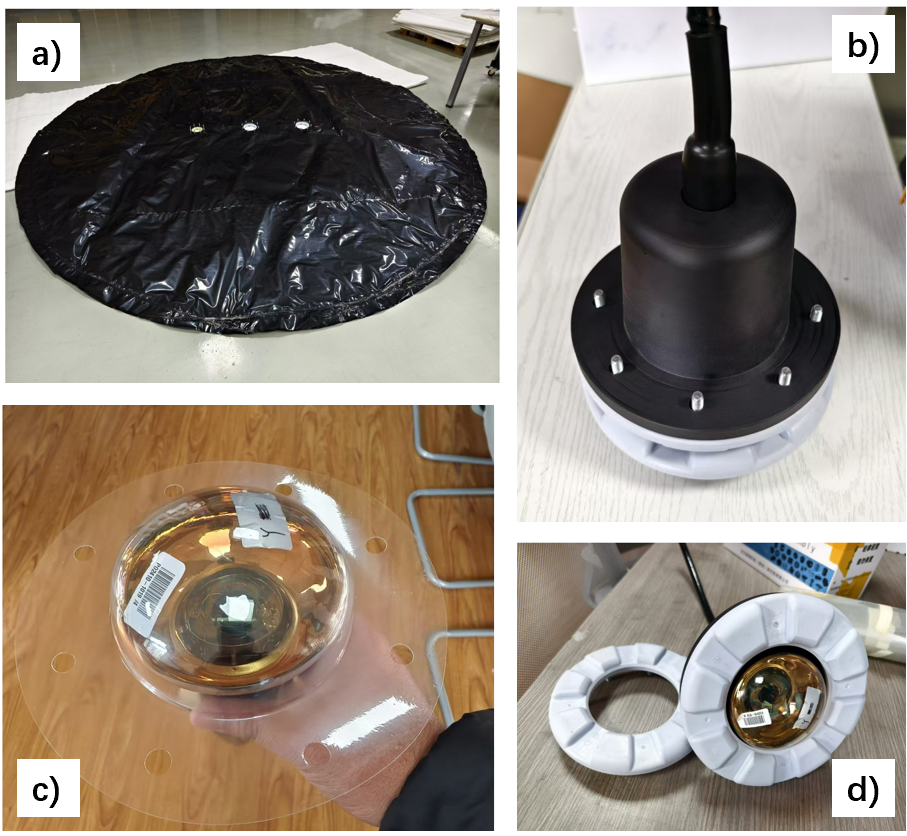}
    \caption{Hardware parts of the prototype detector. a) Double-layered Tyvek® reflective lining, becomes to cylinder along with water filling; b) Encapsulated PMT; c) The contacts between PMT and water is separated by clear plastic cover of 0.2 mm; d) The PMT is coupled with Tyvek lining by flange.}
    \label{fig:hardwares}
\end{figure}

The prototype draws on proven water Cherenkov designs from gamma-ray observatories such as LHAASO  \cite{zhao2014design,zuo2015design,he2018design} and HAWC \cite{deyoung2012hawc}, but incorporates critical adaptations for neutrino detection. The cylindrical tank, with a radius and height of 1.0 m, \xg{was optimized through CORSIKA-based \cite{heck1998corsika} EAS simulations and GEANT4-based \cite{agostinelli2003geant4} water tank simulations to achieve a balance between photon yield and cost-efficiency for PeV-scale $\nu_{\tau}$ interactions.} See Fig.~\ref{fig:tank_design}. A 4-inch hemispherical photomultiplier tube (PMT; North Night N2041) is centrally positioned at the tank top, maximizing angular uniformity while minimizing time jitter. To suppress ambient light and ultraviolet (UV) contamination, the tank walls employ black UV-opaque high-density polyethylene (PE, 10 mm thickness) with a double-layer Tyvek® reflective lining (\textgreater98\% diffuse reflectivity at 300–500 nm in water). Unlike gamma-ray detectors such as HAWC, which utilize non-reflective black linings to suppress stray light, this design adopts full-surface reflectivity to enhance single-photon detection efficiency. The tank is filled with ultra-purified water (attenuation length \textgreater50 m at 400 nm) to minimize absorption of Cherenkov photons in the blue-UV spectrum. A modular calibration system is integrated through a light-tight port on the tank lid, housing a 400-nm LED for two critical functions: (1) PMT gain stabilization via pulsed-light measurements, and (2) characterization of time jitter in first-photon detection through controlled single-photon experiments. The hardware parts are shown in Fig.~\ref{fig:hardwares}.

\subsection{Muon telescope system}
\begin{figure}
    \centering
    \includegraphics[width=250pt]{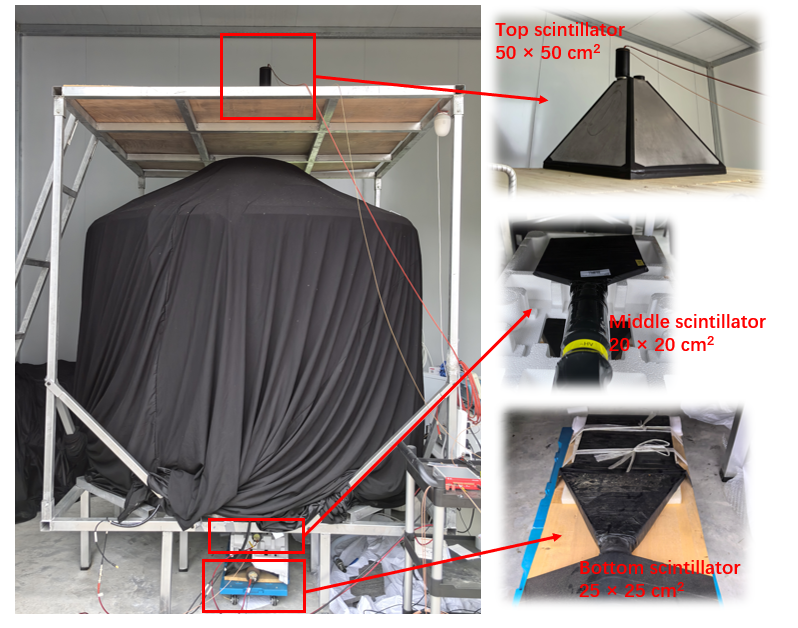}
    \caption{The Cherenkov detector prototype with a three-layer plastic scintillator muon telescope platform with tiered geometry (50×50 cm² at the top, 20×20 cm², 25×25 cm² at the bottom) for cosmic-ray background suppression and detector validation.}
    \label{fig:muon_telescope}
\end{figure}

A muon telescope test platform was developed to evaluate the prototype’s performance and suppress cosmic-ray background. The system utilizes a three-layer plastic scintillator array with a top panel of 50 × 50 cm$^{2}$, a middle panel of 20 × 20 cm$^{2}$, and a bottom panel of 25 × 25 cm$^{2}$. See Fig.~\ref{fig:muon_telescope}. A triple-coincidence logic with a 200 ns timing window is applied to minimize accidental triggers, reducing background noise to negligible levels (\textless10$^{-8}$ Hz, negligible compared to $\sim$0.05 Hz muon event rate). The tiered geometry enhances spatial resolution, enabling precise mapping of the prototype’s detection uniformity across its active area. Each scintillator layer is instrumented with a PMT to generate trigger signals. The measured trigger rates of each scintillator align with the expected sea-level muon flux, confirming the system’s reliability for validating detector response and identifying efficiency variations under real-world operating conditions.

\subsection{Simulations}
Both CORSIKA and GEANT4 simulations were adopted to guide the detector design and validate its performance. CORSIKA simulations of $\nu_{\tau}$-EAS (1–100 PeV) revealed that secondary particles at ground level are dominated by MeV-scale gamma rays (\textgreater80\% of total particles). Complementary Geant4 simulations incorporated the prototype’s exact geometry, including the reflective lining, water attenuation, and PMT response (including quantum efficiency and collection efficiency). The model was calibrated using muon telescope data, achieving \textless5\% discrepancy between simulated and measured signals. By replacing muons with gamma sources in the validated Geant4 framework, the detector’s gamma sensitivity was extrapolated, yielding a \textgreater75\% detection efficiency \xg{(average from multiple injection locations and angles, without weighting)} for 10 MeV–100MeV gammas. See more details in Section 5. This simulation framework bridges neutrino shower physics with detector response, enabling reliable predictions for neutrino signals.

\subsection{PMT selection}
The 4-inch hemispherical PMT was chosen to optimize photon collection uniformity and efficiency across the tank’s active volume. Benefiting from the full-reflective inner lining, which mitigates magnetic field inhomogeneity effects by redistributing photon paths, the PMT operates without additional magnetic shielding. A gain of 4×10$^{6}$ balances two requirements: resolving single-muon signals and detecting 100 PeV neutrino-induced showers without saturation. CORSIKA simulations confirm that the maximum cathode current from such events remains below 5 nA, well within the PMT’s 10 nA cathode saturation threshold. This ensures linear response across the target energy range while maintaining signal integrity under extreme illumination conditions.

\section{Experiment setup}
The experimental setup utilized a suite of high-energy particle detection modules manufactured by CAEN to ensure precise signal acquisition and processing. The setup includes a DT5533M high-voltage power supply, an N978 amplifier, an N843 discriminator with low jitter output, an N455 coincidence logic unit with a programmable time window, an N1145 counter, a custom analog signal splitter and a DT5742 waveform digitizer with 12-bit resolution and 5 GS/s sampling rate, enabling nanosecond-scale timing analysis of Cherenkov photons. 

\section{Water Cherenkov prototype detector performance for MIP particles}

\subsection{MIP signal}
\begin{figure}
    \centering
    \includegraphics[width=200pt]{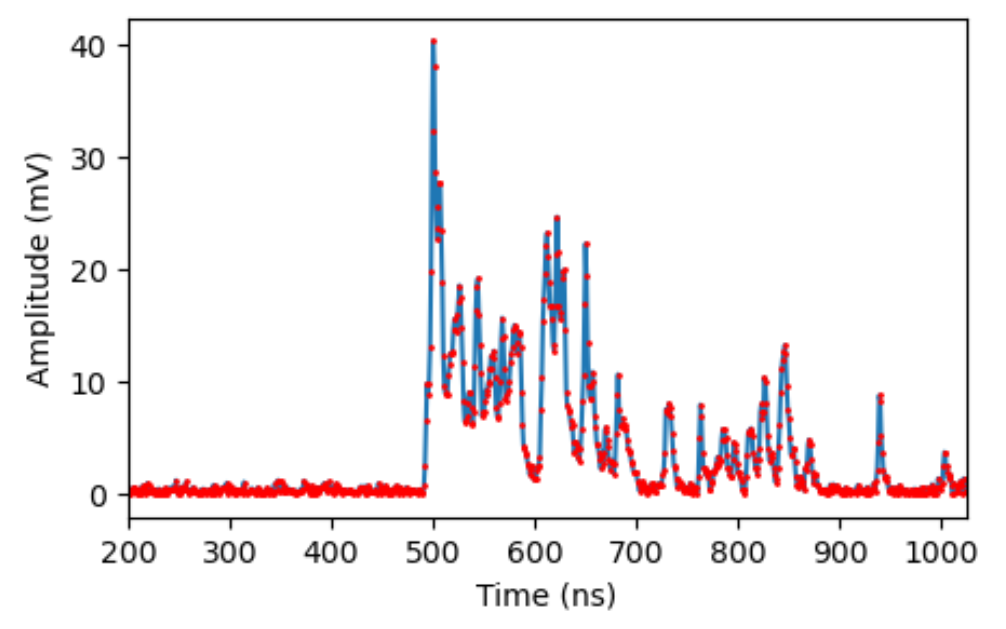}
    \caption{Example waveform of a single muon-induced Cherenkov signal acquired by the DAQ system, showing sharp rising edge (\textless10 ns) and exponential decay ($\sim$600 ns).}
    \label{fig:muon_wave}
\end{figure}

\begin{figure}
    \centering
    \includegraphics[width=200pt]{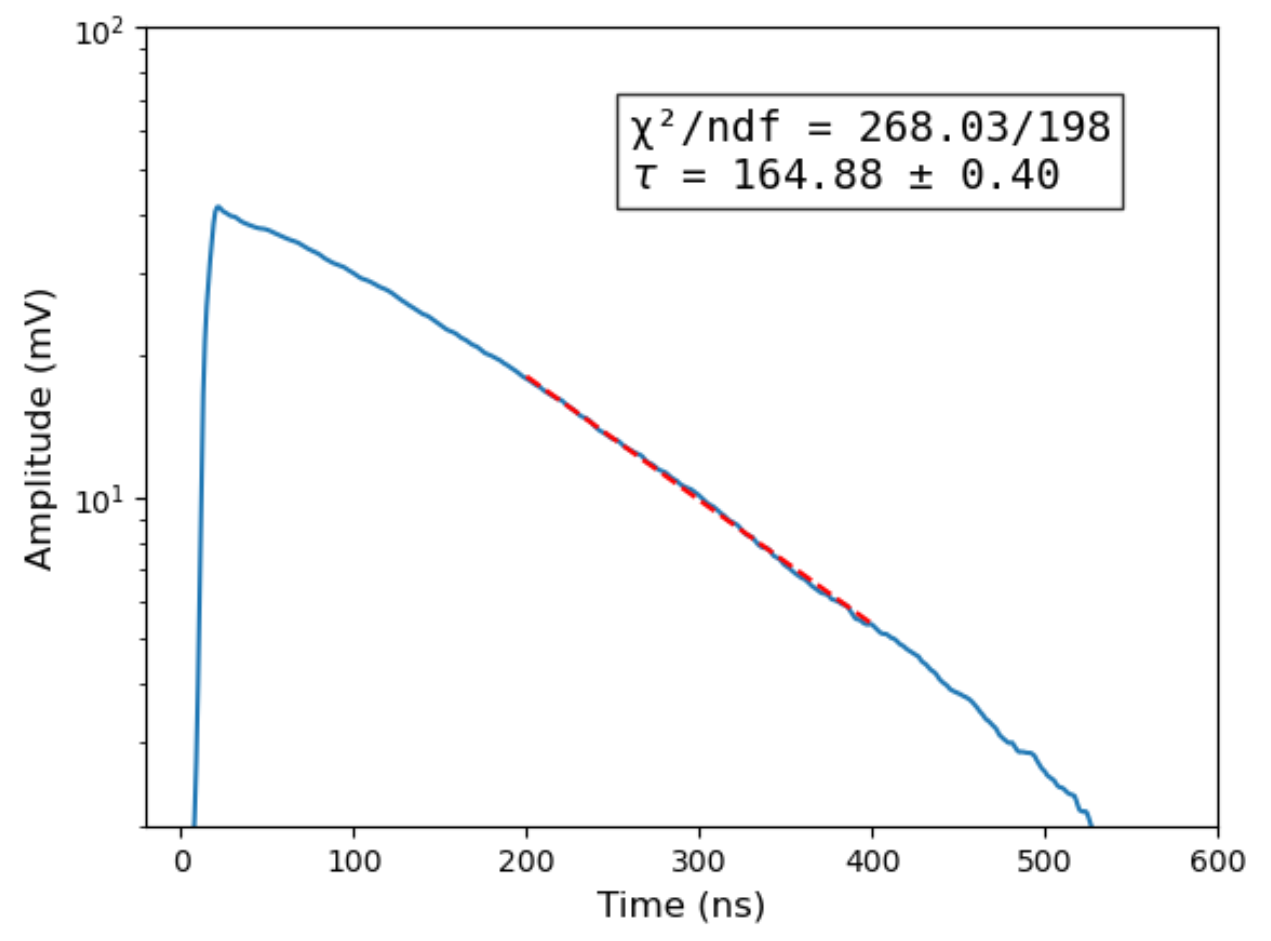}
    \caption{Average waveform of 10,000 muon events after baseline subtraction, demonstrating stable signal-to-noise ratio (SNR \textgreater 15).}
    \label{fig:muon_wave_avr}
\end{figure}

Muon-induced waveforms were captured using the triple-coincidence trigger system and digitized at 1 ns sampling intervals. The signals exhibited a sharp rising edge (\textless10 ns) followed by a long exponential decay ($\sim$600 ns), consistent with Cherenkov photon timing in \xg{such a detector}. Fig.~\ref{fig:muon_wave} shows an example waveform of single muon signals acquired by the DAQ and Fig.~\ref{fig:muon_wave_avr} shows the average waveform acquired over 10,000 samples. Baseline characterization utilized a pre-trigger region of 200 ns, where Gaussian fits revealed a root-mean-square (RMS) noise level below 2 mV and baseline drift under 5 mV over extended operation. The baseline subtraction method, combining pre-trigger averaging and Gaussian fitting, proved robust across varying environmental conditions, achieving a stable SNR \textgreater15. Integrated charge measurements (20–600 p.e.) and peak amplitudes (40–300 mV) demonstrated high repeatability, validating the reliability of the signal processing pipeline for muon calibration and neutrino-induced shower reconstruction.

\begin{figure}
    \centering
    \includegraphics[width=200pt]{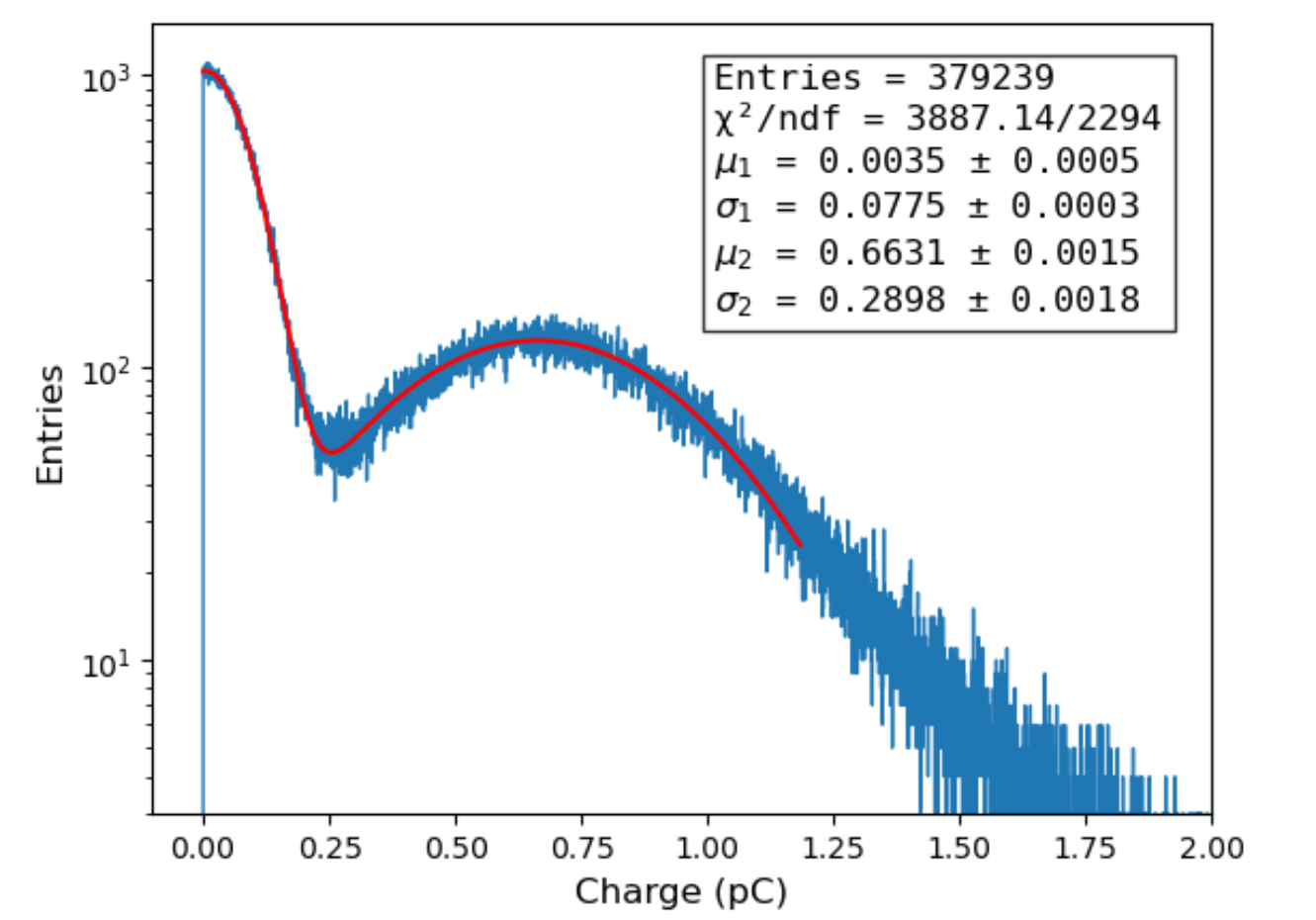}
    \caption{Single-photoelectron (SPE) charge spectrum from 400 nm LED calibration, yielding PMT gain of $4.18\times10^6$ with 0.66 pC per PE.}
    \label{fig:SPE}
\end{figure}

To quantify the muon signal strength, the waveform charge was integrated over a 600 ns window, chosen to span three photon attenuation lengths in ultrapure water, ensuring \textgreater99\% collection of Cherenkov photons. Charge calibration was performed using an in-situ LED system operating in single-photon emission mode (400 nm wavelength). Analysis of the LED’s Poisson-distributed single-photoelectron (SPE) signals yielded a PMT gain of 4.18×10$^{6}$ , corresponding to a single-photon charge of 0.66 pC, as shown in Fig.~\ref{fig:SPE}.

\begin{figure}
    \centering
    \includegraphics[width=200pt]{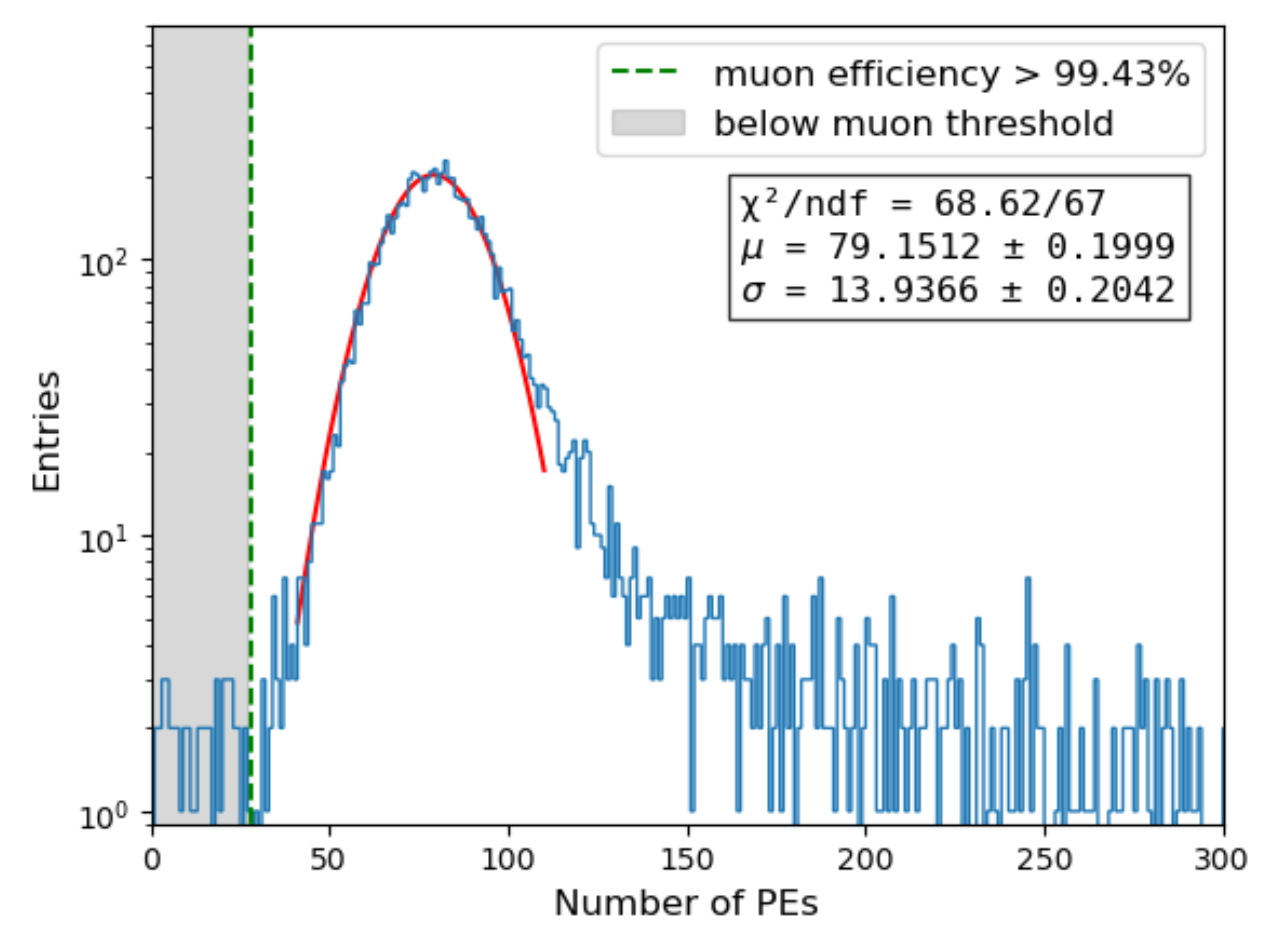}
    \caption{Gaussian-fit muon charge spectrum showing most probable value (MPV) of 79 PE and 17.6\% charge resolution.}
    \label{fig:muon_eff}
\end{figure}

A Gaussian distribution was fitted to the muon charge spectrum, revealing a most probable value (MPV) of 79 PE with a 1$\sigma$ resolution of 13.9 PE, equivalent to a charge resolution of 17.6\%. See Fig.~\ref{fig:muon_eff}. The resolution is dominated by Poisson fluctuations in photon yield ($\sim$12\%) and PMT gain variations ($\sim$5\%), with negligible contributions from electronic noise (\textless1\%). \xg{The position variation} of PE number is less than 5\% (muon injection position from center to edge). The 600 ns integration window and high PMT gain collectively ensure precise reconstruction of muon energy deposition, validating the detector’s capability to resolve charged-particle signals with sub-20\% resolution—a critical requirement for high-energy neutrino experiments.

\subsection{MIP \xg{detection} efficiency}
Quantifying the detector’s muon detection efficiency is critical to validate its capability to resolve charged particles, a prerequisite for reliable neutrino-induced shower reconstruction. Efficiency measurements were derived by comparing the prototype’s trigger rate (signals exceeding a 28 photoelectron (PE) threshold) with the triple-coincidence muon telescope data, which served as a ground truth reference. With $10^{4}$ muon events collected across 72 hours, the prototype detected $\sim$99.5\% of the telescope-tagged muons. The position unevenness of muon detection efficiency is less than 0.5\% (muon injection position changed from center to edge).

The high efficiency \xg{at such a} stringent PE threshold underscores the detector’s exceptional photon collection capability, attributable to the full-reflective tank lining and ultra-purified water medium. This result demonstrates the detector’s robustness in rejecting noise while maintaining near-unity efficiency for minimally ionizing particles, a key requirement for high-energy neutrino observatories operating in high-background environments.

\subsection{Timing resolution}
\begin{figure}
    \centering
    \includegraphics[width=200pt]{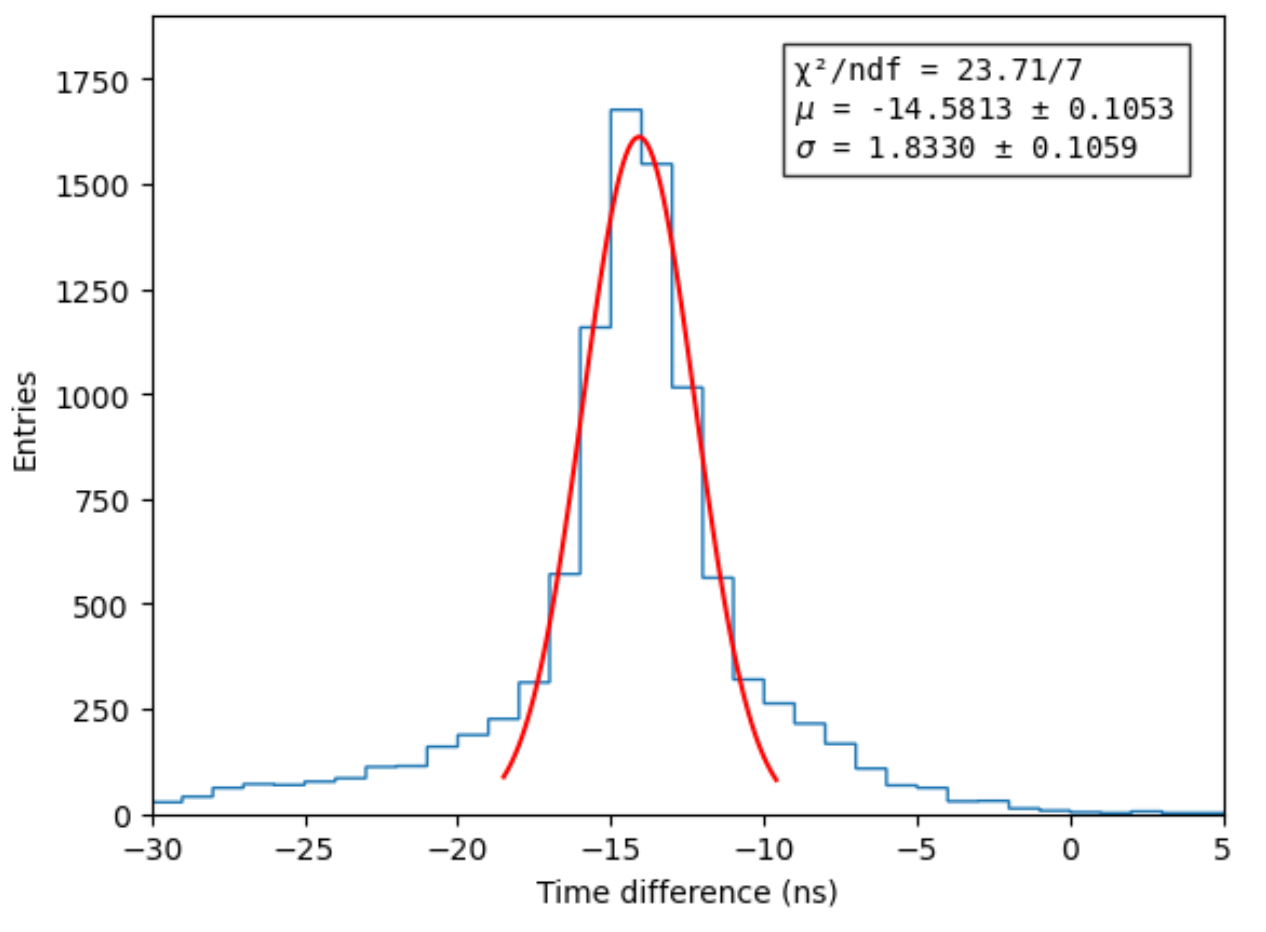}
    \caption{Time difference ($\Delta t$) distribution between muon telescope trigger and PMT signals, indicating intrinsic timing resolution of $\sigma$=1.83±0.1 ns.}
    \label{fig:time_resolution}
\end{figure}

Precise timing resolution is critical for reconstructing the geometry of neutrino-induced air showers, where sub-nanosecond precision is required to resolve Cherenkov photon arrival fronts. To evaluate the prototype’s timing performance, the time difference ($\Delta t$) between signals from the muon telescope’s top scintillator (reference timestamp) and the water Cherenkov detector’s PMT was analyzed. This method isolates systematic delays (e.g., cable lengths, electronics) while attributing the remaining spread to the detector’s intrinsic time resolution. Timestamps were defined at the 20\% amplitude point of the waveform rising edge to minimize walk effects. The measured $\Delta $ distribution Fig.~\ref{fig:time_resolution} showed a Gaussian profile with $\sigma$ = 1.83 ± 0.1 ns, representing the combined resolution of the scintillator and water detector. To estimate the prototype’s standalone resolution, we conservatively assume the scintillator trigger system’s contribution roughly of the value $\sigma_{trg}$ $\approx$ 0.8–1.2 ns. Subtracting this range in quadrature yields an upper limit for the water detector’s intrinsic timing resolution of $\sigma_{intrc}$ = 1.5–1.6 ns.

While further dedicated calibration is needed to disentangle the two components, the total timing resolution of 1.8 ns already satisfies the requirements for reconstructing $\nu_{\tau}$ showers at PeV energies. This performance stems from the PMT's fast rise time (\textless2 ns) combined with the geometry-designed suppression of photon path dispersion. Simulations indicate that a 1.8 ns resolution translates to \textless1° angular uncertainty for 10 PeV neutrino events, sufficient to resolve source directions in a future array configuration.

\subsection{Time jitter analysis}
Understanding the time jitter of the first detected photons is essential for neutrino detectors, as low-energy neutrino interactions (e.g., MeV-scale events) or sub-cascades in $\nu_{\tau}$ showers may produce only a few photons, whose arrival times dominate the reconstruction uncertainty. To quantify this effect, we analyzed LED-generated single-photon events and measured the time delay between the LED trigger and the PMT signal rising edge (50\% threshold crossing).

\begin{figure}
    \centering
    \includegraphics[width=200pt]{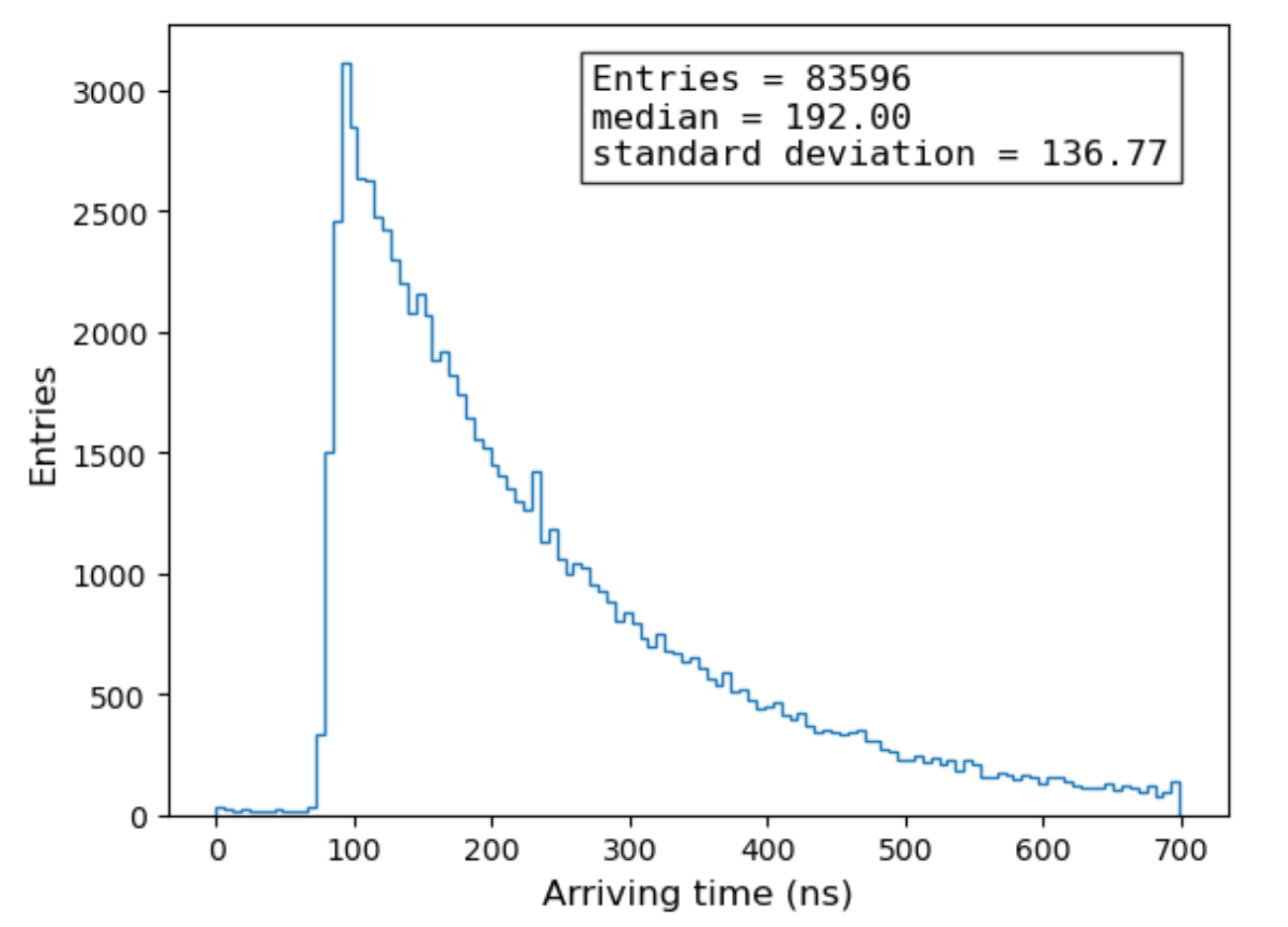}
    \caption{Time jitter distribution of LED-generated single-photon events, dominated by geometric path variations (60–500 ns range).}
    \label{fig:time_jitter}
\end{figure}

The observed jitter spanned \textgreater90\% in 60–500 ns (Fig.~\ref{fig:time_jitter}), predominantly caused by geometric path variations in the 1 m-radius cylindrical detector. Analytic calculation shows even direct-path photons ($\leq$1 reflection) exhibit significant time spreads. \xg{For such a cylindrical detector, the path difference between minimum (center reflection) and maximum (edge-to-edge reflection) trajectories reaches 1.65 m, yielding a 7 ns maximum time spread in water.} While photons undergoing Tyvek multi-reflections exhibit extended delays of 20–800 ns. PMT transit time spread contributed minimally (\textless2 ns), as verified by single-photon calibration. 

For high-energy muons (\textgreater1 GeV) generating \textgreater80 PEs, statistical averaging     suppresses the jitter to \textless2 ns. However, low-energy gammas (\textless100MeV) producing only $\sim$5 PEs suffer from uncorrected jitter, severely degrading timing performance (Note that the measured time jitter of 130 ns of 1 standard deviation under single-PE conditions represents the worst-case scenario; in typical 3-10 PEs of gamma induced events, the timing resolution can improve to 60-90 ns). A practical mitigation is multi-PMT coincidence: requiring coincident timestamps (i.e., \textless50 ns window) across multiple PMTs statistically rejects late-arriving photons. Alternatively however, another approach could be maintaining all data online while performing offline analysis to selectively retain low-jitter events.

This analysis highlights the critical role of geometric effects in small-scale detectors and underscores the need for modular designs in future neutrino observatories to suppress timing uncertainties through spatial sampling.

\section{The GEANT4 model of water Cherenkov detector and its performance on gamma particles}
To bridge experimental measurements with neutrino detection performance predictions, a high-fidelity GEANT4 model was developed to simulate the detector’s optical response. Key parameters—including the Tyvek reflector’s diffuse albedo, water attenuation length, and PMT quantum efficiency - were iteratively refined using muon charge and timing data as constraints. Once calibrated, the validated model was extended to simulate gamma-ray interactions (\textgreater10 MeV), which dominate the secondary particle flux in neutrino-induced showers. This approach leverages the well-understood muon signature to anchor detector physics, enabling reliable extrapolation to low-energy gamma sensitivity critical for neutrino event reconstruction.

\subsection{The propagation of photons in water}
The propagation of photons in water Cherenkov detectors is governed by two primary mechanisms: exponential attenuation due to water absorption and probabilistic attenuation from surface reflections. When charged particles traverse water at superluminal speeds, they generate Cherenkov photons, whose propagation dynamics are characterized as follows:

1. Exponential attenuation due to water absorption

Photon absorption in water follows the Beer-Lambert law. For an initial photon count $N_{0}$, the surviving photons after traveling a distance x are given by:
\begin{equation}
N = N_{0}\times e^{-x/\lambda}\times e^{-x/\lambda_{S}},
\label{eq:absorption}
\end{equation}
where $\lambda$ denotes the water absorption length, representing the average propagation distance before absorption, and $\lambda_{S}$ represents water scattering length. $\lambda$ is highly sensitive to water purity, with typical values ranging from tens to hundreds of meters for ultra-purified water, but is difficult to measure accurately. 

2. Attenuation from surface reflections

The Tyvek inner surface exhibits high diffuse reflectivity ($f\sim$95\%-99\%, varies by wavelength). Each photon impact on the surface has a survival probability determined by the product of reflectivity $f$ and Tyvek coverage ratio $r$. After $n$ reflections, the photon count attenuates as:
\begin{equation}
N = N_0\times(f \times r)^{n}.
\label{eq:reflection}
\end{equation}
The number of reflections $n$ relates to the propagation distance $x$ through the mean step length $L$ (average distance between reflections): $n=x/L$. Thus, reflection-induced attenuation becomes:
\begin{equation}
N = N_0\times(f \times r)^{x/L}.
\label{eq:reflection_attenuation}
\end{equation}

3. Combined attenuation model

Integrating absorption and reflection effects, the total photon attenuation is:
\begin{equation}
N(x) = N_0\times e^{-x/\lambda} \times (f \times r)^{x/L}.
\label{eq:combined_attenuation}
\end{equation}
This model highlights two critical mechanisms: Exponential absorption term $e^{- x/\lambda}$ dominates long-distance attenuation and geometric reflection term $(f\times r)^{x/L}$ separates attenuation governed by detector geometry. GEANT4 simulations of LHAASO reveal that after multiple reflections, photon directions become fully randomized, stabilizing the mean step length $L$ \cite{li2019novel}. For a cylindrical detector (1 m radius, 1 m height), $L$ is about 1.2 m. This stability ensures $L$ remains constant regardless of initial photon direction, validating the geometric dependence in \eqref{eq:combined_attenuation}. Here, the scattering term is neglected, since the observed attenuation remains insensitive to medium scattering effects, as scattered photons remain confined within the tank. In ultra-purified water, scattering contributions become negligible relative to absorption processes, particularly in their impact on the characteristic mean step length of photon propagation.

4. Temporal Distribution Characteristics

The temporal photon distribution $N(t)$ is derived by converting spatial attenuation to time-domain decay via $x=vt$:
\begin{equation}
N(t) = N_0 \times e^{-t/\tau}, \quad \tau = \frac{\lambda_m}{v},
\label{eq:time_decay}
\end{equation}
where $\lambda_{m}$ represents the effective attenuation length, combining water absorption and surface reflection:
\begin{equation}
\frac{1}{\lambda_m} = \frac{1}{\lambda} - \frac{\ln(f) + \ln(r)}{L}.
\label{eq:effective_attenuation}
\end{equation}
This relationship underpins experimental methodologies for measuring optical parameters. 

\subsection{The GEANT4 model}
The GEANT4 simulation framework in this study incorporates wavelength-dependent optical parameters validated through the LHAASO collaboration’s UNIFIED model \cite{born2013principles, wang2012study}, which has been successfully applied to large-scale Cherenkov detector simulations. Key parameters, including the water absorption length and Tyvek reflectivity, are defined \xg{for a range wavelengths} from 295 nm to 620 nm, corresponding to the emission of Cherenkov photons. The absorption length is set to 150 m (at 425 nm) based on LHAASO’s experimental measurements, avoiding redundant long-tube calibration for small-scale detectors, as simulations confirmed insensitivity (\textless5\% variation in PE counts) to absorption length values above 100 m. The Tyvek reflectivity $f$ of 99\% aligns with laboratory characterizations of Tyvek film around 425 nm, while the PMT quantum efficiency (28\% at 375nm) and photon collection efficiency (75\%) are derived from manufacturer specifications. The Tyvek coverage ratio $r$ is about 99\%, considering the losses caused by wrinkling.

\begin{figure}
    \centering
    \includegraphics[width=200pt]{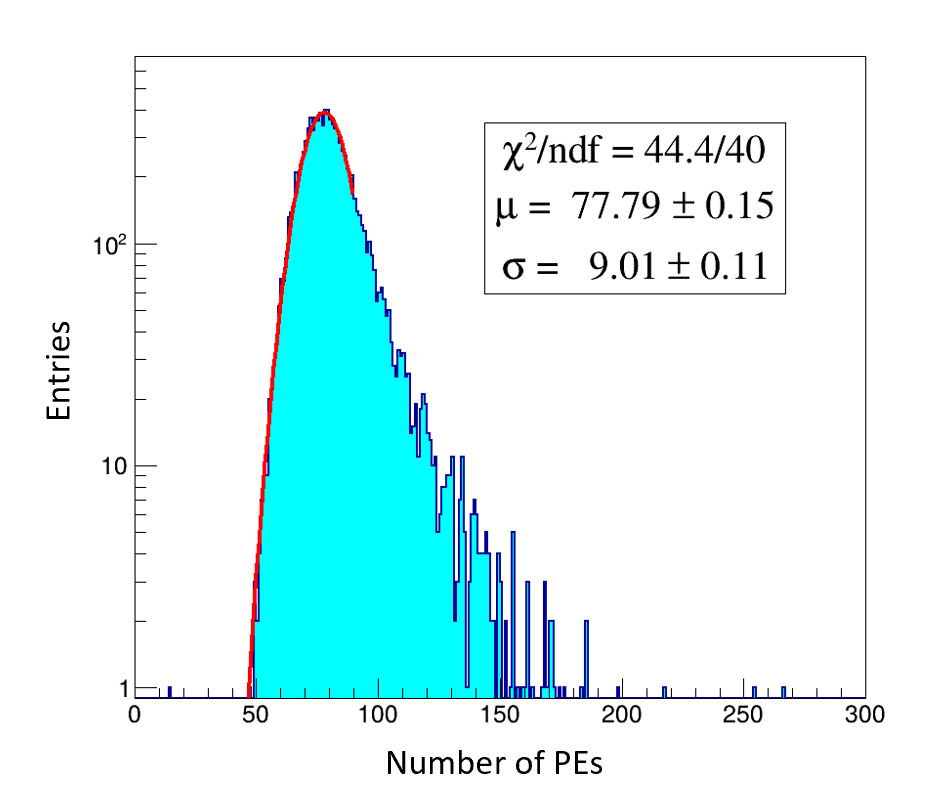}
    \caption{GEANT4-simulated photoelectron (PE) count distribution for 3 GeV muons, matching experimental data within 2\% deviation.}
    \label{fig:G4_PE_num}
\end{figure}

\begin{figure}
    \centering
    \includegraphics[width=200pt]{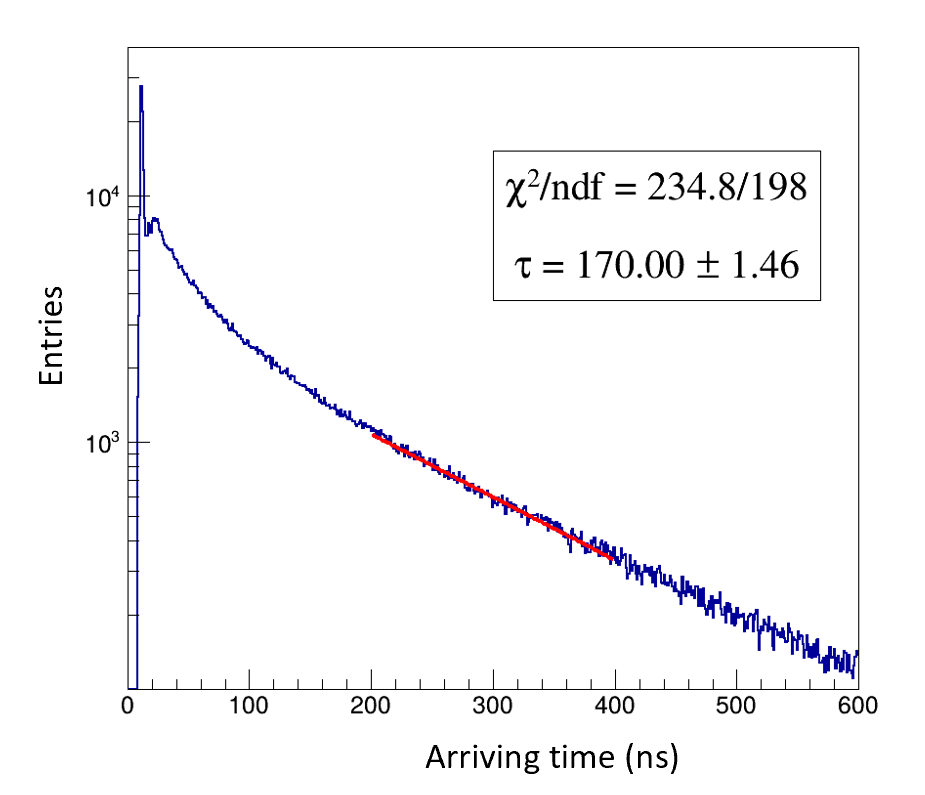}
    \caption{Simulated photon arrival time distribution showing exponential decay ($\tau$=169.5±1.5 ns), consistent with measured 164.9 ns attenuation.
}
    \label{fig:G4_arriving_t}
\end{figure}

The simulation models 3 GeV muons vertically incident at 50 cm from the detector center, generating Cherenkov photons whose propagation is tracked until detection or absorption. Photoelectron (PE) numbers and arrival time distributions are extracted, revealing a mean simulated PE count of 77.8 ± 9 (Fig.~\ref{fig:G4_PE_num}) and a decay time constant of 170.0 ± 1.5 ns (Fig.~\ref{fig:G4_arriving_t}). These results closely match experimental data from muon telescope measurements (79 ± 14 PE and 164.9 ± 0.4 ns), with deviations below 5\%, confirming the validity of the optical parameterization. The effective attenuation length derived from the time constant ($\lambda_{m}$ = 38.1 m) further agrees with theoretical predictions based on the combined absorption-reflection model.

Discrepancies between simulation and experiment arise from idealized assumptions in the GEANT4 framework. The simulation employs monoenergetic muons with point-like incidence, whereas real-world data involve a broad cosmic-ray muon energy spectrum (1–10 GeV) and spatial dispersion (±15 cm) due to the telescope’s angular acceptance. These factors introduce PE number fluctuations and timing spread. Additionally, the simulation omits electronic response effects—such as PMT transit time spread and amplifier bandwidth limitations—which smear experimental waveforms, optimistically estimating the timing resolution. Environmental noise, including dark counts and ambient light leakage, further broadens the baseline noise in measurements (2 mV RMS) compared to noise-free simulations.

Future refinements can be improved by integrating realistic muon energy spectra generated by the CRY cosmic-ray library \cite{hagmann2012cosmic} and incorporate distributed incidence patterns within the telescope’s acceptance. Digitization effects, including measured PMT impulse responses and amplifier saturation, will be embedded to bridge the gap between optical photon tracking and electronic signal outputs. The UNIFIED model’s proven scalability ensures its applicability to both current small-scale prototypes and future large-scale neutrino detectors, providing a robust foundation for optimizing detection efficiency and background rejection in next-generation observatories.

\subsection{The detector performance on low-energy gamma sub-particles}
\begin{figure}
    \centering
    \includegraphics[width=200pt]{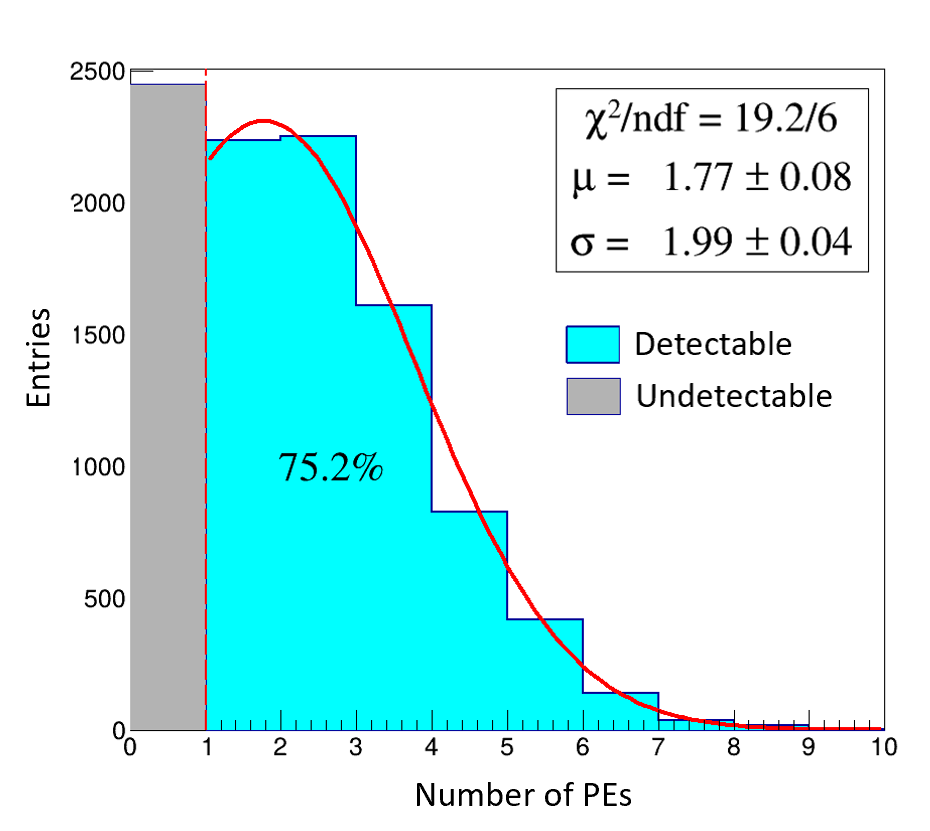}
    \caption{Detection efficiency for 10 MeV gamma rays in the validated GEANT4 model, achieving \textgreater75\% efficiency despite Cherenkov threshold limitations.}
    \label{fig:gamma_10_eff}
\end{figure}

\begin{figure}
    \centering
    \includegraphics[width=200pt]{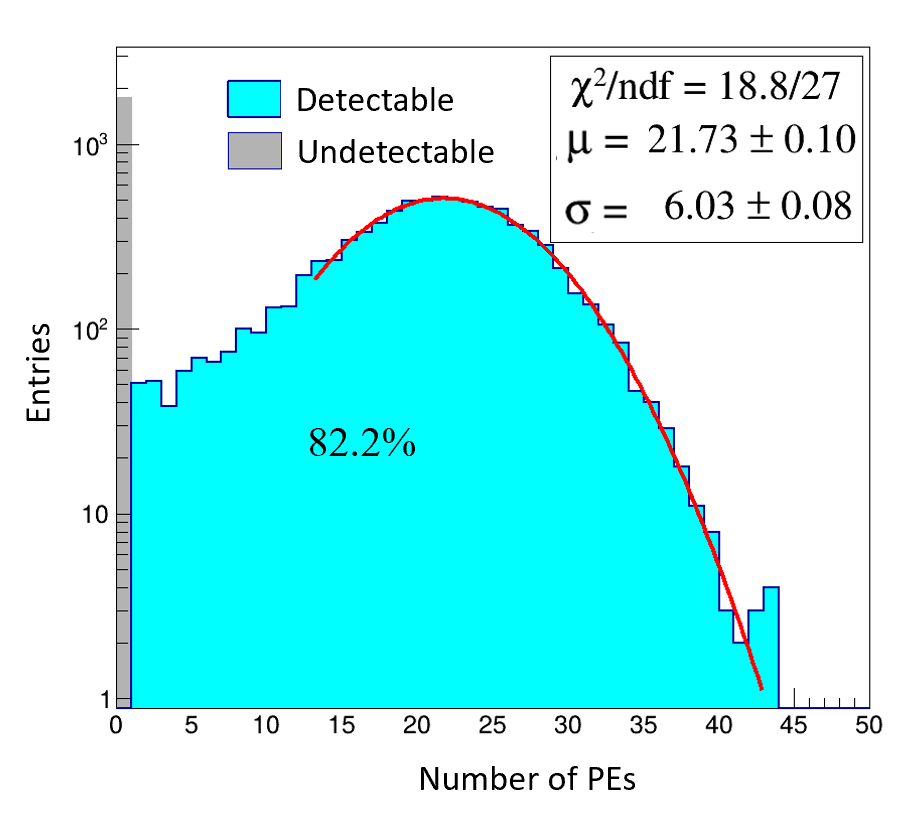}
    \caption{Detection efficiency for 100 MeV gamma-ray simulations, demonstrating 22 PE/event yield with \textgreater80\% detection efficiency.}
    \label{fig:gamma_100_eff}
\end{figure}

\begin{figure}
    \centering
    \includegraphics[width=200pt]{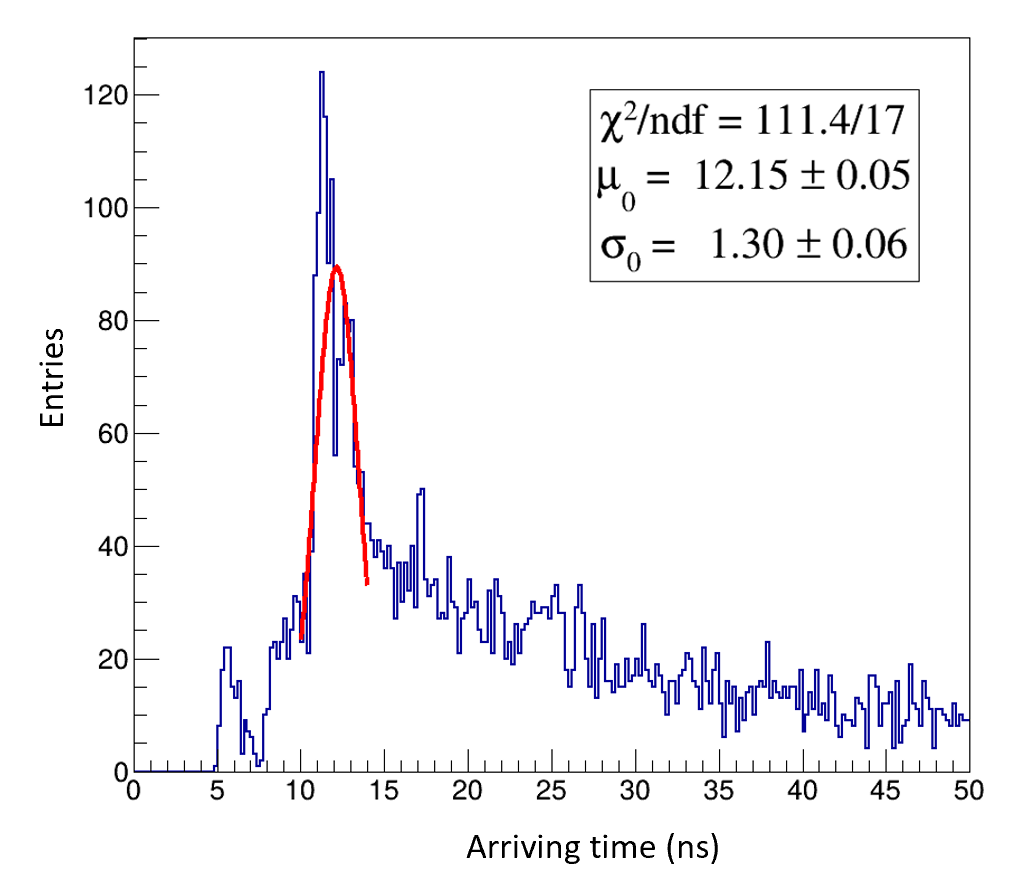}
    \caption{Timing resolution for 10 MeV gamma events, showing sub-2 ns resolution despite 40 ns geometric jitter.}
    \label{fig:gamma_10_TR}
\end{figure}

The detector's response to low-energy gamma rays was systematically evaluated using the validated GEANT4 model, with particular focus on 10 MeV and 100 MeV gamma rays vertically incident at 50 cm from the detector center. These energy ranges are particularly relevant for studying secondary gamma photons produced in $\nu_{\tau}$ interactions within the 1-100 PeV energy range. Simulation results demonstrated that 10 MeV gamma rays generated an average of 1-4 PEs per event with a detection efficiency exceeding 75\% (Fig.~\ref{fig:gamma_10_eff}), while 100 MeV gamma rays produced approximately 22 PEs per event at efficiencies above 80\% (Fig.~\ref{fig:gamma_100_eff}). \xg{The nonlinearity between PE yields and energy is governed by threshold effects (pair production at 1.022 MeV and Cherenkov emission at 0.26 MeV) alongside energy deposition mechanisms in water.} The timing resolution for 10 MeV gammas is below 2 ns while the time jitter spans to 40 ns (Fig.~\ref{fig:gamma_10_TR}).

The observed detection efficiencies, while lower than the near-unity efficiency for muons, represent a significant improvement over conventional gamma detection technologies. For comparison, plastic scintillator systems coupled with lead converters typically achieve 5-15\% detection efficiency. The performance difference stems from several fundamental physical processes and detector characteristics. First, gamma rays must be above 1.022 MeV of pair production threshold and generate the electron-position pair for Cherenkov photon emission. For 10 MeV gamma rays in water, the pair production length is about $\frac{9}{7}X_0 \approx 46$ cm \cite{kunwar2023double}, where $X_0 \approx 36$ cm is the radiation length in water.  This means roughly 12\% of incident 10 MeV gamma rays traverse the 1 m detector without interaction.

Furthermore, the secondary electrons and positrons produced must exceed the Cherenkov threshold kinetic energy of 0.26 MeV in water to emit detectable photons. While 100 MeV gamma rays typically produce high-energy pairs averaging 30 MeV, the pairs from 10 MeV gamma rays average only 3 MeV, with about 1/3 failing to emit effective Cherenkov light. \xg{Compton scattering is another channel transferring energy to recoil electrons that may fall below the Cherenkov emission threshold around 10 MeV.} The limited number of photons produced, particularly for 10 MeV events where only several Cherenkov photons are generated on average, leads to significant statistical fluctuations. The detector geometry also impacts performance. Cherenkov photons are emitted in a characteristic conical pattern with an opening angle of approximately 41° relative to the particle trajectory. For near-vertical gamma showers in small detectors, a significant fraction of photons may miss the PMT entirely. 

Despite these limitations, the water Cherenkov technique demonstrates superior performance compared to plastic scintillator systems by a factor of 15-20 in efficiency. This advantage arises from water's higher density and superior gamma conversion rates compared to plastic scintillators, without requiring additional heavy-metal converters. Additionally, the directional nature of Cherenkov emission enables more precise reconstruction of interaction vertices compared to the isotropic light emission characteristic of scintillation detectors.

The simulation results confirm the detector's capability to resolve neutrino-induced electromagnetic showers, with 100 MeV gamma events providing sufficient signal-to-noise ratio (S/N \textgreater 10) for reliable triggering. These findings establish a foundation for understanding the detector's neutrino detection thresholds and background rejection capabilities. The demonstrated sensitivity to low-energy gamma rays, particularly in the 10-100 MeV range, positions this detector design as a valuable tool for studying neutrino interactions and related phenomena. More detailed discussion about the detection ability of $\nu_{\tau}$-induced EAS will be in the next session.

\section{Discussion: Detector viability for $\nu_{\tau}$-induced EAS detection}

\begin{figure}
    \centering
    \includegraphics[width=200pt]{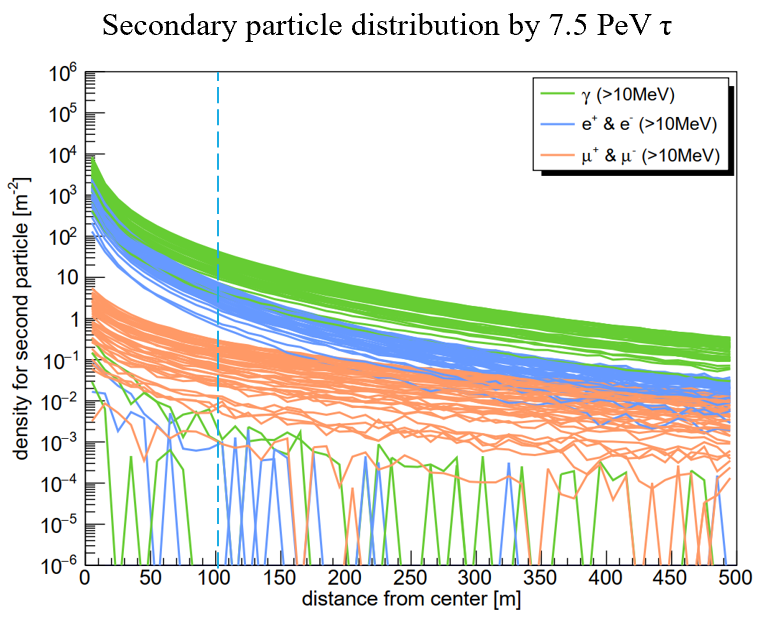}
    \caption{CORSIKA-simulated secondary particle density from 7.5 PeV $\nu_{\tau}$-induced air showers, dominated by \textgreater10 MeV gamma rays at 100 m core distance.}
    \label{fig:tau_7PeV}
\end{figure}

The proposed water Cherenkov detector design demonstrates foundational capabilities for identifying secondary particles from $\nu_{\tau}$-induced extensive air showers (EAS) based on CORSIKA simulations of 7.5 PeV neutrinos. At 100 m from the shower core, simulations predict the particle density to be 10–100 gamma rays (\textgreater10 MeV), \xg{1–10 electrons/positions (\textgreater10 MeV), and 0.01–1 muons per m$^2$}, as shown in Fig.~\ref{fig:tau_7PeV}. Therefore, the particle density and the detection efficiency (see Section 5.3) are high enough to make a $\geq$2 PE triggering threshold per station experimentally achievable. These results establish the unit detector's intrinsic capacity to resolve EAS electromagnetic components assuming a 100 m array grid, though the system-level efficiency requires optimization of array topology and readout architecture beyond the current scope.

To adapt this unit detector design for array deployment, several requirements can be reasonably foreseen. The intrinsic fluctuations of extensive air showers (EAS)-particularly in neutrino interaction depths and secondary particle multiplicities—introduce significant uncertainties, necessitating effective rejection of accidental triggers such as cosmic-ray muons or hadronic showers. Consequently, optimizing triggering efficiency and background suppression becomes critical for array performance. Implementing a multi-detector coincidence strategy emerges as a necessary design for the array. This requires spatially correlated triggers (e.g., $\geq$5 adjacent units) firing within optimized timing windows to suppress random backgrounds to negligible levels while preserving neutrino signal efficiency. The timing window optimization must be systematically investigated to maximize the SNR. Besides, sub-degree angular resolution requires rigorous characterization of timing performance and synchronization. \xg{This necessitates constructing a prototype array with dedicated readout electronics and White Rabbit systems, coupled with development of advanced angle reconstruction algorithms.} Further guidance on refining these performance metrics requires detailed array-scale simulations and experimental validation. Critically, while these design challenges extend beyond the scope of unit-level validation, they remain fundamentally constrained by the prototype’s capabilities, which does not preclude a successful array implementation.

The modularity of the detector allows scalable deployment across square-kilometer arrays, with cost-effectiveness driven by passive water shielding and simplified electronics. \xg{However, the configuration of individual detector units significantly impacts detection efficiency and timing performance. For instance, PMT positioning and density favor direct Cherenkov photon detection from specific incident angles without requiring reflections-as implemented in HAWC with bottom-mounted PMTs. Combined with its non-reflective enclosure, avoiding photon time tailing, this setup achieves exceptional time resolution. Conversely, fully reflective liners (e.g., LHAASO KM2A) maximize photon collection efficiency but introduce severe time tailing, necessitating advanced event reconstruction algorithms (e.g., rapid identification of earliest-arriving photons). Partial reflector designs emerge as a motivated compromise between these performance metrics. For canyon-based neutrino arrays, the optimal configuration must balance array-level performance against cost-effectiveness. This requires detailed simulations and experiments, comparing individual detector designs to identify the most viable solution.}

Beyond neutrino detection, this design can be easily adapted to gamma-ray detector array by adopting a ``double-layer" detector variant, as proposed by \cite{kunwar2023double}. \xg{The upper electromagnetic layer serves as a gamma detector and absorber, while the bottom layer operates as a muon detector. To enhance gamma absorption, a lead inter-layer is positioned between two independently encapsulated sections, without direct water contact while capturing residual penetrating gammas.} \xg{This approach eliminates the need for costly underground muon detectors while preserving full $\nu_{\tau}$ sensitivity and achieving gamma-ray detection capabilities comparable to established observatories like HAWC.}
 
\section{Conclusion}
This study pioneers a canyon-adapted water Cherenkov detector for $\nu_{\tau}$ arrays, offering a viable and scalable solution for ground-based $\nu_{\tau}$ detection. Laboratory validation confirms exceptional performance: \textgreater99\% MIP detection efficiency with minimal spatial non-uniformity (\textless0.5\%) and \textless2 ns timing resolution, enabling precise air shower reconstruction. Simulations further demonstrate robust gamma sensitivity (\textgreater75\% at 10 MeV), critical for resolving electromagnetic sub-cascades in $\nu_{\tau}$-induced EAS. The prototype’s simplified modular design, leveraging natural canyon shielding, enables a more cost-efficient deployment with respect to conventional arrays. These advancements establish this design as a dual-purpose platform for next-generation neutrino-gamma observatories, positioning it to probe the PeV–EeV frontier through neutrino and gamma-ray measurements. This study addresses critical sensitivity gaps in current neutrino detectors and establishes canyon-optimized water Cherenkov arrays as the foundational technology for next-generation $\nu_{\tau}$ astronomy.

\section*{Declaration of competing interest}
The authors declare that they have no known competing financial interests or personal relationships that could have appeared to influence the work reported in this paper.

\section*{Data availability statement}
Data will be made available on request.

\section*{Acknowledgments}
We sincerely thank Huihai He and Yiqing Guo from the Institute of High Energy Physics for their expert guidance and constructive feedback. We are grateful to Yizhong Fan, Jianhua Guo and Xiaoyuan Huang at Purple Mountain Observatory for their valuable supports and insights. We also appreciate the contributions of graduate students Huan Yang and Junjie Zhai. Special thanks to staffs from Qingdao Hengxin Plastic Co., Ltd. for their essential support to the experiments. This research used resources from the Project for Young Scientists in Basic Research of the Chinese Academy of Sciences (No. YSBR-061).

\bibliographystyle{elsarticle-num} 
\bibliography{reference}

\end{document}